\documentclass[twocolumn,english,aps, prl, amsmath, amssymb]{revtex4-1}
\usepackage[T1]{fontenc}
\usepackage[latin9]{inputenc}
\usepackage[letterpaper]{geometry}
\usepackage{verbatim}
\usepackage{amsbsy}
\usepackage{amssymb}
\usepackage{graphicx}
\usepackage{esint}

\makeatletter
\@ifundefined{textcolor}{}
{%
 \definecolor{BLACK}{gray}{0}
 \definecolor{WHITE}{gray}{1}
 \definecolor{RED}{rgb}{1,0,0}
 \definecolor{GREEN}{rgb}{0,1,0}
 \definecolor{BLUE}{rgb}{0,0,1}
 \definecolor{CYAN}{cmyk}{1,0,0,0}
 \definecolor{MAGENTA}{cmyk}{0,1,0,0}
 \definecolor{YELLOW}{cmyk}{0,0,1,0}
}

\makeatother

\usepackage{babel}
\begin{document}

\title{Tunable circular dichroism due to the chiral anomaly in Weyl semimetals}

\author{Pavan Hosur}

\affiliation{Department of Physics, Stanford University, Stanford, CA 94305, USA}

\author{Xiao-Liang Qi}

\affiliation{Department of Physics, Stanford University, Stanford, CA 94305, USA}
\begin{abstract}
Weyl semimetals are a three dimensional gapless topological phase
in which bands intersect at arbitrary points -- the Weyl nodes --
in the Brillouin zone. These points carry a topological quantum number
known as the \emph{chirality} and always appear in pairs of opposite
chiralities. The notion of chirality leads to anomalous non-conservation
of chiral charge, known as the \emph{chiral anomaly}, according to
which charge can be pumped between Weyl nodes of opposite chiralities
by an electromagnetic field with non-zero $\boldsymbol{E}\cdot\boldsymbol{B}$.
Here, we propose probing the chiral anomaly by measuring the optical
activity of Weyl semimetals via circular dichroism. In particular,
we observe that applying such an electromagnetic field on this state
gives it a non-zero gyrotropic coefficient or a Hall-like conductivity,
which may be detectable by routine circular dichroism experiments.
This method also serves as a diagnostic tool to discriminate between
Weyl and Dirac semimetals; the latter will give a null result. More
generally, any experiment that probes a bulk correlation function
that has the same symmetries as the gyrotropic coefficient can detect
the chiral anomaly as well as differentiate between Dirac and Weyl
semimetals.
\end{abstract}
\maketitle

\section{Introduction}

Weyl semimetals (WSMs) are a novel gapless topological phase of matter
and are accruing considerable attention of late\cite{HosurWeylReview,KrempaWeyl,VafekDiracReview,TurnerTopPhases}.
They are three-dimensional systems whose band structure contains isolated
points in momentum space where a pair of non-degenerate bands intersect.
These intersection points -- or Weyl nodes -- can be assigned a \emph{handedness}
or \emph{chirality} $\chi=\pm1$; very general arguments show that
the total number of Weyl nodes in the Brillouin zone must be even
with half of each chirality\cite{NielsenFermionDoubling1,NielsenFermionDoubling2}.
Moreover, they are topological in the sense that they can only be
annihilated (gapped out) in pairs of opposite chirality, or via superconductivity.
Thus, they are stable as long as translational symmetry and charge
conservation hold. Near the Weyl nodes, the dispersion is linear and
the Hamiltonian resembles the Hamiltonian for Weyl fermions well-known
in high-energy physics:
\begin{equation}
H_{W}=\chi\boldsymbol{k}\cdot\boldsymbol{\sigma}
\end{equation}
where $\boldsymbol{k}$ is the momentum relative to the Weyl node
and $\sigma_{i}$ are Pauli matrices in the local band basis. Hence,
the name WSMs.

WSMs are bestowed with a physical property known as the Adler-Bell-Jackiw
anomaly or the chiral anomaly\cite{NielsenABJ,AjiABJAnomaly,BasarTriangleAnomaly,SonSpivakWeylAnomaly,ZyuninBurkovWeylTheta,LandsteinerAnomaly,GoswamiFieldTheory,HosurWeylTransport}.
This is a well-known phenomenon in high energy physics. It represents
an anomalous non-conservation of chiral charge in the presence of
appropriate external electromagnetic fields, even though the Hamiltonian
enjoys the continuous symmetry -- the chiral gauge symmetry -- that
is expected to lead to chiral charge conservation via Noether's theorem.
The resolution to this paradox lies in the fact that the chiral gauge
transformation modifies the integration measure in the path integral,
and hence the path integral itself. Thus, the chiral anomaly is a
purely quantum process that the classical Hamiltonian is oblivious
to. Viewed differently, it is an artifact of the low energy theory,
and appropriate regularization at high energies that smoothly interpolates
between Weyl Hamiltonians of opposite chiralities would destroy the
chiral gauge symmetry\cite{HosurWeylReview}.

WSMs present a condensed matter realization of the phenomenon. In
this context, the anomaly implies that although the total charge in
the WSM is conserved, the charge in the momentum states near the left-handed
or the right-handed Weyl nodes is not individually conserved. In the
simplest case of a WSM with just two Weyl nodes, the anomaly can be
written as
\begin{equation}
\partial_{\mu}j_{\mbox{ch}}^{\mu}=\frac{e^{2}}{4\pi^{2}\hbar^{2}}\boldsymbol{E}\cdot\boldsymbol{B}\label{eq:anomaly equation}
\end{equation}
where $j_{\mbox{ch}}^{\mu}=(j_{+}^{\mu}-j_{-}^{\mu})/2$ is the four
dimensional chiral current and the subscripts $\pm$ on the currents
denote the chirality of the Weyl node contributing to them, and the
right hand side states that the pumping is driven by an electromagnetic
field configuration with non-zero $\boldsymbol{E}\cdot\boldsymbol{B}$
\footnote{Henceforth, we abbreviate ``electromagnetic fields with non-zero
$\boldsymbol{E}\cdot\boldsymbol{B}$'' as ``an $\boldsymbol{E}\cdot\boldsymbol{B}$
field'' or parallel electric and magnetic fields%
}. The electromagnetic fields are space and time-dependent in general.
In the absence of any spatial variations, (\ref{eq:anomaly equation})
reduces to 
\begin{equation}
\partial_{t}\rho_{\mbox{ch}}=\frac{e^{2}}{4\pi^{2}\hbar^{2}}\boldsymbol{E}(t)\cdot\boldsymbol{B}(t)\label{eq:charge-pumping}
\end{equation}
For time-independent $\boldsymbol{E}$ and $\boldsymbol{B}$, $\rho_{\mbox{ch}}$
grows linearly with time until a scattering process cuts off the growth
by relaxing the charge imbalance between the Weyl nodes. Such processes
are rare in clean systems since they involve a large momentum transfer.
Thus, any measurement of the chiral anomaly can unambiguously distinguish
between WSMs and the less exotic Dirac semimetals, in which Weyl nodes
of opposite chiralities coincide in momentum and energy. (\ref{eq:anomaly equation})
is a two-dimensions higher version of the chiral anomaly present at
the edge of an integer quantum Hall state\cite{Maeda2DAnomaly}. There,
charge can be pumped from one edge to the other by a longitudinal
electric field: $\partial_{\mu}j_{\mbox{ch}}^{\mu}=\frac{e^{2}}{2\pi\hbar}E$.
An important difference, however, is that the chiral currents $j_{\pm}^{\mu}$
in the integer quantum Hall state reside on spatially separated edges
and can be observed individually with local probes, whereas the chiral
currents in WSMs are separated in momentum space and cannot be distinguished
by spatially local probes. The question is, what kind of probe can
qualitatively see the chiral anomaly? 

A transport phenomenon intimately tied to the chiral anomaly that
was noticed early on was a large longitudinal magnetoconductivity.
This occurs because relaxation of chiral charge involves large momentum
scattering and hence takes a long time in clean systems\cite{NielsenABJ}.
Recently, another transport experiment was proposed in which chiral
charge pumping according to (\ref{eq:charge-pumping}) resulted in
a large enhancement of the length scale over which an applied local
voltage decayed\cite{ParameswaranNonLocalTransport}. However, these
effects are quantitative rather than qualitative and hence, difficult
to identify unambiguously in magnetotransport data\cite{BiSbKimMagnetoTransportExpt},
especially if the intervalley relaxation time is short. While the
first effect is accompanied by weak localization or weak antilocalization
physics\cite{Garate2012} in $\mathcal{T}$-symmetric WSMs and would
have to be isolated from it, the enhancement of the length scale in
the second proposal may not be large enough to be measurable for moderate
intervalley relaxation rates because the non-local voltage falls exponentially
with the rate. There exist various other predictions for chiral transport
phenomena, most famously, a non-quantized anomalous Hall effect\cite{ChenAxionResponse,FangChernSemimetal,GoswamiFieldTheory,RanQHWeyl,VafekDiracReview,WeylMultiLayer,ZyuninBurkovWeylTheta}
and the\emph{ chiral magnetic effect}\cite{ZhouSemiclassicalTransport,VazifehEMResponse,BasarTriangleAnomaly,ChenAxionResponse},
in which a current flows along an applied magnetic field. An optical
signature of the anomalous Hall effect, namely, anomalous birefringence,
was briefly discussed in Ref \onlinecite{GrushinWeyl}. However,
a more fundamental question remains unanswered, namely, what kind
of material properties are sensitive to the chirality of a given system?
Experiments that measure these properties can, in principle, be designed
to \emph{qualitatively }rather than \emph{quantitatively}, probe the
chiral anomaly in WSMs as well.

\section{Gyrotropy}

In this work, we propose that material parameters or transport coefficients
that are of the form of time-reversal-invariant ($\mathcal{T}$-invariant)
rank-3 pseudotensors are appropriate probes of the anomaly. Like the
chirality of a system, these quantities are odd under inversion ($\mathcal{I}$)
and even under $\mathcal{T}$ and hence, directly couple to it. They
can therefore be employed to distinguish between Weyl nodes of opposite
chirality and consequently probe the chiral anomaly. Below, we elaborate
on one such material property -- one which is responsible for optical
activity. However, the fact that $\mathcal{T}$-invariant rank-3 pseudotensors
exist only in chiral systems is more general and can be used to probe
the chiral anomaly in other kinds of experiments as well. For instance,
a chiral strain field, such as that present near a screw dislocation,
will modify the resistivity by an amount that depends on the handedness
of the dislocation as well as to the chirality of the underlying band
structure. The relevant tensor for this phenomenon is the elastoresistive
tensor, which describes the change in resistivity due to applied strain.
On the other hand, these tensors, despite being non-vanishing in general
in all chiral systems, are not tunable in ordinary chiral systems
such as sugar molecules. It is only in WSMs that the chiral anomaly
can be exploited to tune the magnitude and sign of chiral transport.

The $\mathcal{T}$-invariant rank-3 pseudotensor that is responsible
for optical activity is the gyrotropic tensor $\gamma_{ijk}$, which
is defined in terms of the dielectric tensor $\epsilon_{ij}(\boldsymbol{q},\omega)$
as\cite{Landau1984}
\begin{equation}
\epsilon_{ij}(\boldsymbol{q},\omega)=\epsilon_{ij}^{0}(\omega)+i\gamma_{ijk}(\omega)q_{k}+\mathcal{O}(q^{2})\,\,\,.\label{eq:gamma-def}
\end{equation}
where $\boldsymbol{q}$ and $\omega$ are the wavevector and frequency
of light. Clearly, $\gamma_{ijk}$ represents the response to variations
in the electric field. For systems with cubic or higher symmetry,
such as a single isotropic Weyl node, this tensor is purely anti-symmetric:
$\gamma_{ijk}=\gamma\varepsilon_{ijk}$, and the gyrotropic coefficient
reduces to a complex number $\gamma$. Similarly, the $q$-independent
diagonal part of the dielectric tensor is proportional to the Kronecker
delta function: $\epsilon_{ij}^{0}(\omega)=\epsilon_{0}(\omega)\delta_{ij}$
for a single isotropic Weyl node. Isotropy of the Weyl node can be
assumed without losing any essential physics, since any anisotropy
can be removed by rescaling momenta around the node.

$\mathcal{T}$-symmetry leads to Onsager's reciprocity condition:
$\epsilon_{ij}(\boldsymbol{q},\omega)=\epsilon_{ji}(-\boldsymbol{q},\omega)$,
which allows a non-zero $\gamma$, while mirror symmetries allow only
even powers of momentum normal to the mirror plane in the dielectric
tensor. Thus, $\gamma$ vanishes in systems that have a mirror symmetry.
Systems that break all mirror symmetries and hence, break $\mathcal{I}$-symmetry,
have a non-vanishing $\gamma$ in general and can be assigned a handedness
proportional to $\gamma$. In particular, a single Weyl node is chiral
and can have a non-zero $\gamma\propto\chi$. However, any symmetries
relating Weyl nodes of opposite chiralities will make the total $\gamma$
of the WSM vanish. Since spatially local probes only detect the total
response of all the Weyl nodes, they will then see a null result for
$\gamma$. To get a non-zero result, one must find a way to subtract,
rather than add, contributions to $\gamma$ from Weyl nodes of opposite
chirality.

Such a subtraction can be done by observing that $\gamma$ is related
to the Hall conductivity through the relation $\epsilon_{ij}=\delta_{ij}+\frac{i\sigma_{ij}}{\varepsilon_{0}\omega}$,
where $\varepsilon_{0}$ is the permittivity of free space, so it
must have contributions that are odd in the charge of the quasiparticles.
If one can somehow arrange for the doping to be different at the $\chi=+1$
and $\chi=-1$ Weyl nodes, their contributions to $\gamma$ will no
longer cancel. The anomaly induced by $\boldsymbol{E}\cdot\boldsymbol{B}$
((\ref{eq:anomaly equation}) and (\ref{eq:charge-pumping})) precisely
ensures such a charge imbalance between the two chiralities. In other
words, once a finite amount of charge has been pumped from $\chi=-1$
to $\chi=+1$, the $\chi=-1$ ($\chi=+1$) Weyl node is surrounded
by a hole (electron) Fermi surface assuming they were undoped initially.
If they were already doped, their local Fermi levels will become different.
In the language of symmetries, chiral charge pumping ensures that
all symmetries relating Weyl nodes of opposite chiralities are broken
because they have different Fermi levels relative to the Weyl points.

That systems with non-zero $\gamma$ exhibit circular dichroism --
conversion of linearly polarized light into elliptically polarization
as it propagates through the system -- can be seen as follows. The
eigenmodes of the dielectric tensor (\ref{eq:gamma-def}) correspond
to circularly polarized light; the associated eigenvalues determine
the refractive indices for the two polarizations, $n_{L,R}^{2}(\omega)=\epsilon_{0}(\omega)\pm\gamma q$.
Later, we will find that $\gamma$ for the system of interest is purely
imaginary. Since $\mathcal{I}m(n_{L})\neq\mathcal{I}m(n_{R})$, the
two circular polarizations making up the linear polarization are absorbed
by different amounts as the light propagates through the system, resulting
in circular dichroism. It is straightforward to show that the ellipticity
of the light that comes out is given by 
\begin{equation}
\left|\tan\theta_{CD}\right|\equiv\left|\frac{E_{R}-E_{L}}{E_{R}+E_{L}}\right|\approx\frac{\left|\mathcal{I}m\left[\gamma\right]\right|\ell\omega^{2}}{2c^{2}}\label{eq:CD-angle}
\end{equation}
for $\left|\gamma\omega/c\right|\ll\epsilon_{0}$, where $E_{L,R}$
are the transmitted amplitudes of the circularly polarized fields,
$\ell$ is the thickness of the sample and $c$ is the speed of light.
Thus, we propose detecting the chiral anomaly \emph{optically}. Later
in the paper, we estimate the sizes of this effect for typical WSMs
and find it may be within experimental limits.

A caveat, though, is that WSMs can at most preserve only one out of
$\mathcal{I}$ and $\mathcal{T}$ symmetries. Consequently, a $\mathcal{T}$-symmetric
WSM already has a non-zero $\gamma$ in general, without external
fields, while an $\mathcal{I}$-symmetric WSM has vanishing $\gamma$
but can be optically active due to possible ferromagnetic moments
that break $\mathcal{T}$. Later we will describe how the background
contributions to optical activity can be separated from the anomaly-based
ones by a clever separation of their frequencies.

\begin{figure}
\begin{centering}
\includegraphics[height=1.9cm]{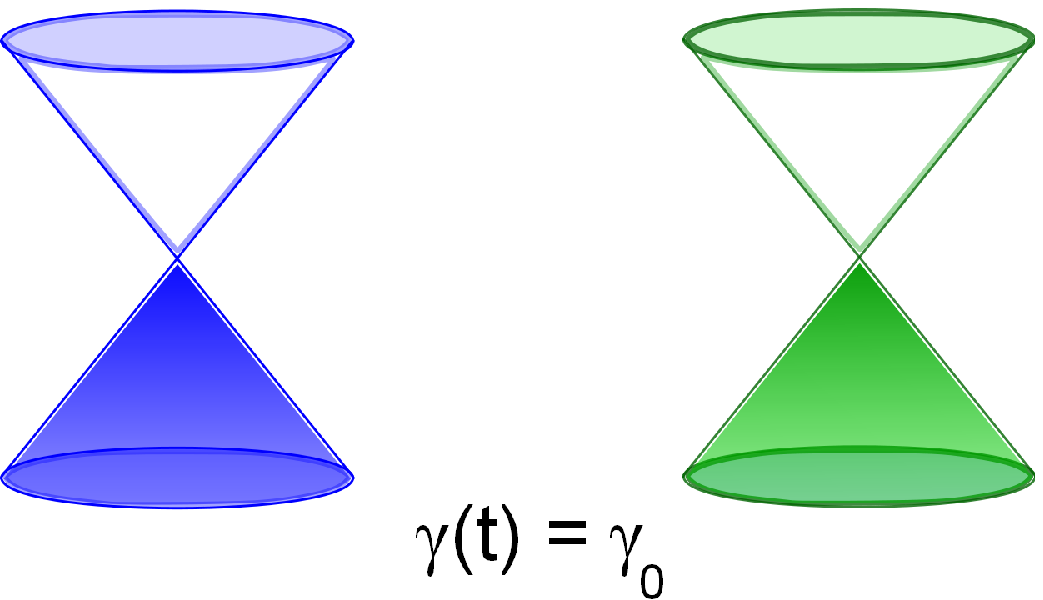}~~~\includegraphics[height=2cm]{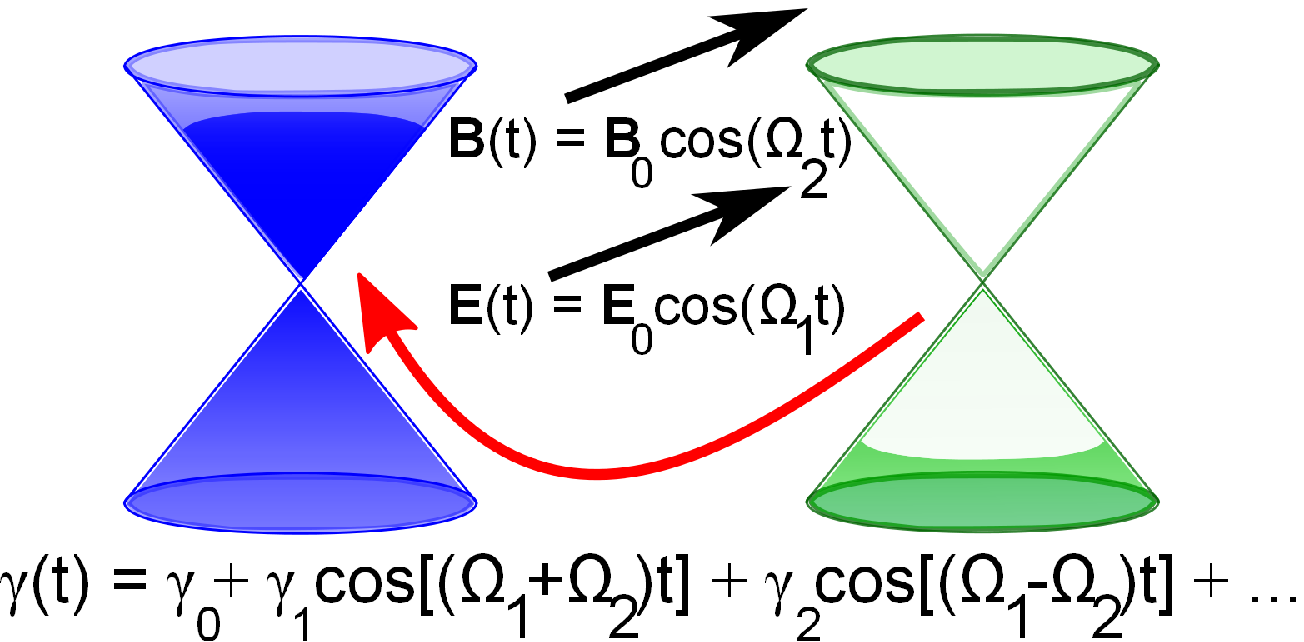}
\par\end{centering}

\caption{Illustration of gyrotropy induced by an $\boldsymbol{E}\cdot\boldsymbol{B}$
field in WSMs. Colored (white) regions denote filled (empty) states
and the two colors indicate Weyl nodes of opposite chiralities, which
are located at different points in momentum space. In the absence
of $\boldsymbol{E}\cdot\boldsymbol{B}$, the Fermi levels at the two
nodes are equal. An $\boldsymbol{E}\cdot\boldsymbol{B}$ field pumps
charge across the nodes, shown by the curved red arrow, resulting
in a charge imbalance between the nodes and a consequent net anomalous
contribution to $\gamma$. Since WSMs break $\mathcal{I}$ or $\mathcal{T}$,
they are in general optically active in the absence of external fields
as well. The anomalous contribution, proportional to $\boldsymbol{E\cdot B}$,
can be isolated by applying electromagnetic fields at distinct finite
frequencies.}
\end{figure}

\section{Single Weyl node results}

In the following, we will first calculate the gyrotropic coefficient
of a single Weyl node and then apply the results to WSMs with multiple
Weyl nodes. We start with the Hamiltonian $H_{\boldsymbol{k}}^{\chi}=\chi\hbar v_{F}\boldsymbol{k}\cdot\boldsymbol{\sigma}-\mu_{\chi}$
for a single Weyl node $W^{\chi}$ of chirality $\chi$ and chemical
potential $\mu_{\chi}$ above the Weyl point. Such a description is
valid if we equilibration within a Weyl node occurs at time-scales
that are much shorter than the frequency at which the experiment is
performed: $\omega\ll\tau_{intra}^{-1}$. $\mu_{\chi}$ consists of
two parts in general -- the background chemical potential due to doping
already present in the system, and the change because of charge pumping.
For a constant $\boldsymbol{E}\cdot\boldsymbol{B}$ field, which is
what we assume for now, the latter grows linearly with time in the
absence of large momentum scattering; in practice, such scattering
\emph{is} present and the system reaches a non-equilibrium steady
state characterized by an inter-valley relaxation time $\tau_{inter}$.
(\ref{eq:charge-pumping}) dictates that the density of electrons
pumped into the neighborhood of $W_{\chi}$ in this time is $\Delta\rho_{\chi}=\chi\frac{e^{2}}{4\pi^{2}\hbar^{2}}\boldsymbol{E}\cdot\boldsymbol{B}\tau_{inter}$.
Later we argue that time-dependent fields greatly facilitate separating
the anomalous contribution to $\gamma$ from possible background terms,
and modify the results accordingly.

In the low frequency limit, $\omega\tau_{intra}\ll1$, $\gamma_{ijk}$
was shown to be related to the first moment of the Berry curvature
of the occupied states and is thus, particularly easy to calculate\cite{Moore_Orenstein_Berry_Curvature}.
We recap the derivation below following a similar spirit as Ref \cite{Moore_Orenstein_Berry_Curvature},
keeping the application to a single Weyl node in mind.

The semiclassical equations of motion for a wavepacket in a band with
dispersion $\varepsilon(\boldsymbol{k})$ and Berry curvature $\boldsymbol{F}(\boldsymbol{k})=i\boldsymbol{\nabla_{k}}\times\left\langle \psi_{\boldsymbol{k}}|\boldsymbol{\nabla_{k}}\psi_{\boldsymbol{k}}\right\rangle $,
where $|\psi_{\boldsymbol{k}}\rangle$ is the Bloch wavefunction,
in the presence of a space-time varying electric field $\boldsymbol{E}(\boldsymbol{r},t)=\boldsymbol{E}e^{i(\boldsymbol{q}\cdot\boldsymbol{r}-\omega t)}$
read 
\begin{eqnarray}
\dot{\boldsymbol{r}} & = & \boldsymbol{v}(\boldsymbol{k})-\boldsymbol{F}(\boldsymbol{k})\times\frac{e}{\hbar}\boldsymbol{E}(\boldsymbol{r},t)\nonumber \\
\dot{\boldsymbol{k}} & = & \frac{e}{\hbar}\boldsymbol{E}(\boldsymbol{r},t)\label{eq:semi-classical}
\end{eqnarray}
where $\boldsymbol{v}(\boldsymbol{k})=\frac{1}{\hbar}\boldsymbol{\nabla_{k}}\varepsilon(\boldsymbol{k})$.
The second term above describes the anomalous Hall current\cite{HaldaneAHE};
integrating over $\boldsymbol{k}$ gives a Hall current at $\boldsymbol{r}$
due to the local electric field at $\boldsymbol{r}$. However, gyrotropy
stems from a Hall-like response driven by spatial variations in the
electric field. Such a response is non-local, and requires solving
(\ref{eq:semi-classical}) for finite times. Doing so iteratively
to first order in $\boldsymbol{E}$ yields
\begin{eqnarray}
\boldsymbol{r}(t) & = & \boldsymbol{r}_{0}+\boldsymbol{v}(\boldsymbol{k}_{0})t\nonumber \\
 &  & +\intop_{0}^{t}\mathrm{d}t^{\prime}\intop_{0}^{t^{\prime}}\mathrm{d}t^{\prime\prime}\frac{e}{\hbar}\boldsymbol{E}\left(\boldsymbol{r}_{0}+\boldsymbol{v}(\boldsymbol{k}_{0})t^{\prime\prime},t^{\prime\prime}\right)\cdot\boldsymbol{\nabla}_{\boldsymbol{k}_{0}}\boldsymbol{v}(\boldsymbol{k}_{0})\nonumber \\
 &  & -\frac{e}{\hbar}\boldsymbol{F}_{\boldsymbol{k}_{0}}\times\intop_{0}^{t}\mathrm{d}t^{\prime}\boldsymbol{E}\left(\boldsymbol{r}_{0}+\boldsymbol{v}(\boldsymbol{k}_{0})t^{\prime},t^{\prime}\right)\nonumber \\
\boldsymbol{k}(t) & = & \boldsymbol{k}_{0}+\intop_{0}^{t}\mathrm{d}t^{\prime}\frac{e}{\hbar}\boldsymbol{E}\left(\boldsymbol{r}_{0}+\boldsymbol{v}(\boldsymbol{k}_{0})t^{\prime},t^{\prime}\right)
\end{eqnarray}
where $\boldsymbol{r}_{0}$ and $\boldsymbol{k}_{0}$ are the initial
position and wavevector of the wavepacket. The Hall response is given
by the third line of the expression for $\boldsymbol{r}(t)$. Explicitly,
it is
\begin{eqnarray}
\boldsymbol{j}_{\mbox{Hall}}(\boldsymbol{k}_{0},t) & = & e\dot{\boldsymbol{r}}_{\mbox{Hall}}(\boldsymbol{k}_{0},t)\nonumber \\
 & = & \boldsymbol{E}(\boldsymbol{r},t)e^{i\boldsymbol{q}\cdot\boldsymbol{v}_{\boldsymbol{k}_{0}}t}\times\frac{e^{2}}{\hbar}\boldsymbol{F}_{\boldsymbol{k}_{0}}
\end{eqnarray}
Integrating over $\boldsymbol{k}_{0}$ for all occupied states gives
the total Hall current density
\begin{equation}
\boldsymbol{J}_{\mbox{Hall}}(\boldsymbol{r},t)\approx\frac{e^{2}}{\hbar}\boldsymbol{E}(\boldsymbol{r},t)\times\intop_{\boldsymbol{k}\in\mbox{occ}}\boldsymbol{F}_{\boldsymbol{k}}(1+i\boldsymbol{q}\cdot\boldsymbol{v}_{\boldsymbol{k}}t)
\end{equation}
for $|\boldsymbol{q}\cdot\boldsymbol{v_{k}}t|\ll1$. For a single
Weyl node doped away from the Weyl point, the first term vanishes
because of an effective time-reversal symmetry in $H_{\chi}$: $\boldsymbol{k}\to-\boldsymbol{k}$,
$\boldsymbol{\sigma}\to-\boldsymbol{\sigma}$ which results in $\boldsymbol{F_{k}}=-\boldsymbol{F}_{-\boldsymbol{k}}$.
The non-vanishing part of the $\gamma$, proportional to the $\boldsymbol{q}$-linear
part of the Hall conductivity, is
\begin{equation}
\gamma_{single}(\omega)=\frac{e^{2}}{\hbar}\frac{i\tau_{intra}}{\varepsilon_{0}\omega}\intop_{\boldsymbol{k}\in\mbox{occ}}\left(\boldsymbol{F_{k}}\cdot\hat{\boldsymbol{q}}\right)\left(\boldsymbol{v_{k}}\cdot\hat{\boldsymbol{q}}\right)\label{eq:gamma-general}
\end{equation}
In writing (\ref{eq:gamma-general}), we have replaced $t$ by $\tau_{intra}$
to suggest that the linear growth with time is cut-off by a relaxation
process at $t\simeq\tau_{intra}$. The physical meaning of the above
result is as follows. Imagine a pair of electrons, with momentum $\boldsymbol{k}$
and $-\boldsymbol{k}$, starting at the same point in space and traveling
in opposite directions, along $\boldsymbol{q}$ and along $-\boldsymbol{q}$.
They can travel for a time $\tau_{intra}$ without getting scattered;
in the process, they are acted upon by slightly different electric
fields since the electric field varies in space. Since the Berry curvatures
they experience are equal and opposite, their total contribution to
the Hall current is non-vanishing and proportional to the wavevector
corresponding the electric field variations.%
{} For a single Weyl node given by the Hamiltonian $H_{\boldsymbol{k}}^{\chi}=\chi\hbar v_{F}\boldsymbol{k}\cdot\boldsymbol{\sigma}-\mu_{\chi}$
it is straightforward to show that $\boldsymbol{F_{k}}=\chi\frac{\hat{\boldsymbol{k}}}{2k^{2}}$
and $\boldsymbol{v_{k}}=v_{F}\hat{\boldsymbol{k}}$. Substituting
in (\ref{eq:gamma-general}) gives
\begin{equation}
\gamma_{single}(\omega)=i\frac{\chi\mu_{\chi}e^{2}\tau_{intra}}{6\pi^{2}\varepsilon_{0}\hbar^{2}\omega}\label{eq:gamma-single}
\end{equation}
 Note that $\gamma_{single}$ is purely imaginary, so it leads to
circular dichroism.

\section{Applying to real WSM}

Having derived $\gamma$ for a single Weyl node, we now make the leap
to a real WSM, which contains an even number of Weyl nodes with half
of each chirality. Thus, we sum the contributions of all the Weyl
nodes and estimate the size of the consequent gyrotropic circular
dichroism in a real WSM. For simplicity, we assume an $\mathcal{I}$-symmetric
system with two Weyl nodes, one of each chirality. The results can
be trivially scaled to $N$ pairs of nodes by simply multiplying by
$N$. If the Fermi level is tuned to the Weyl point in the absence
of external electromagnetic fields -- the iridate (candidate) WSMs
Y$_{2}$Ir$_{2}$O$_{7}$ and Eu$_{2}$Ir$_{2}$O$_{7}$ are expected
to be in this limit because of their stoichiometry -- then, there
is no background chemical potential and $\mu_{\chi}$ only depends
on the charge pumped:
\begin{equation}
\mu_{\chi}=\chi\left(\frac{3e^{2}\hbar v_{F}^{3}}{2}\boldsymbol{E}\cdot\boldsymbol{B}\tau_{inter}\right)^{1/3}
\end{equation}
Summing (\ref{eq:gamma-single}) for two nodes gives
\begin{equation}
\gamma(\omega)=i\frac{e^{2}\tau_{intra}}{3\pi^{2}\varepsilon_{0}\hbar^{2}\omega}\left(\frac{3e^{2}\hbar v_{F}^{3}}{2}\boldsymbol{E}\cdot\boldsymbol{B}\tau_{inter}\right)^{1/3}\label{eq:gamma-undoped}
\end{equation}
and
\begin{equation}
\left|\tan\theta_{CD}\right|=\frac{\alpha}{3\pi}(\omega\tau_{inter})\frac{\left|\mu_{+}-\mu_{-}\right|\ell}{\hbar c}
\end{equation}
where $\alpha=e^{2}/4\pi\varepsilon_{0}\hbar c\approx1/137$ is the
fine structure constant. Putting in realistic values of parameters,
$v_{F}\approx10^{6}\mathrm{ms^{-1}}$ from the band structure of candidate
WSMs Y$_{2}$Ir$_{2}$O$_{7}$, Eu$_{2}$Ir$_{2}$O$_{7}$\cite{PyrochloreWeyl}
and Cd$_{3}$As$_{2}$\cite{WangCd3As2}, $|\boldsymbol{E}|=10\mathrm{V/mm}$,
$|\boldsymbol{B}|=1\mathrm{T}$, $\tau_{intra}=1\mathrm{fs}$, $\tau_{inter}=100\mathrm{ps}$,
$\hbar\omega=100\mathrm{meV}$, $\ell\sim1\mathrm{\mu m}$ gives $|\mu_{\chi}|\sim10\mathrm{meV}$
and $\left|\tan\theta_{CD}\right|\sim10\mathrm{\mu rad}$. We have
chosen numbers so that $\omega$ exceeds the plasma frequency $\omega_{p}\simeq\sqrt{(2\alpha/3\pi)(c/v_{F})}|\mu_{\chi}|$\cite{SarmaDiracPlasmons,WeylDielectric};
this ensures that the incident light is not screened. While the above
estimate is crude, it suggests that the effect may be measurable by
current experiments. Moreover, the effect will be enhanced in a sufficiently
clean system since both the lifetimes $\tau_{inter}$ and $\tau_{intra}$
will be longer.

On the other hand, if the WSM is doped to a Fermi level of $\epsilon_{F}$
away from the Weyl nodes, the effect is suppressed by $\mathcal{O}\left[\left(\frac{\mu_{+}-\mu_{-}}{\epsilon_{F}}\right)^{2}\right]$
for the same amount of charge pumped because of the density of states
for a 3D linear dispersion is proportional to $\epsilon^{2}$. Moreover,
choosing a frequency that lies between $\tau_{intra}^{-1}$ and $\omega_{p}$
will be difficult and perhaps impossible. Thus, we only focus of the
case where the WSM is undoped to begin with.

\section{Subtracting the background}

Circular dichroism measurements are commonly used to study systems
that break $\mathcal{T}$ or $\mathcal{I}$ symmetry. WSMs break at
least one of $\mathcal{T}$ and $\mathcal{I}$ symmetries and thus,
exhibit intrinsic, i.e., $\boldsymbol{E}\cdot\boldsymbol{B}$-independent
optical activity in general. The final piece of the puzzle of probing
the anomaly via gyrotropy is being able to subtract this background.

One way to do so is to simply do an experiment without $\boldsymbol{E}$
and $\boldsymbol{B}$ fields and subtract the results from the results
in the presence of an $\boldsymbol{E}\cdot\boldsymbol{B}$ field.
While this procedure can work in principle, it involves subtracting
a potentially large background and is thus error-prone. Moreover,
the $\boldsymbol{E}$ and $\boldsymbol{B}$ fields can change the
optical activity independently of the chiral anomaly as well, for
instance, by inducing polarization or magnetization.

A cleaner procedure would be to make the fields time-dependent. Thus,
is one applies $\boldsymbol{E}(t)=\boldsymbol{E}\cos\Omega_{1}t$
and $\boldsymbol{B}(t)=\boldsymbol{B}\cos\Omega_{2}t$, such that
$\Omega_{1,2}\ll\tau_{inter}^{-1}$, and measures the optical activity
at a higher frequency $\omega\gg\tau_{inter}^{-1}$, then $\boldsymbol{E}(t)$
and $\boldsymbol{B}(t)$ can be treated quasistatically and the preceding
analysis can be applied with minor modifications. In particular, the
gyrotropic coefficient will pick up a slow time-dependence: 
\begin{equation}
\gamma(\omega;t)=\frac{4i\alpha c\tau_{intra}}{3\pi\hbar\omega}\left(\frac{3e^{2}\hbar v_{F}^{3}\tau_{inter}}{2}\boldsymbol{E}\cdot\boldsymbol{B}\cos\Omega_{1}t\cos\Omega_{2}t\right)^{1/3}
\end{equation}
for $\tau_{inter}\ll t\ll\Omega_{1,2}^{-1}$, and will thus have components
at frequencies $\Omega_{1}\pm\Omega_{2}$ which should be easily separable
from other frequency components. In addition to the time-dependence
of $\gamma$, its dependence on the relative angle between $\boldsymbol{E}$
and $\boldsymbol{B}$ should be easy to observe as well on top of
the constant background.

\section{Conclusions}

In summary, we have described a method to probe the chiral anomaly
in WSMs optically. Our method is based on the fact that an $\boldsymbol{E}\cdot\boldsymbol{B}$
electromagnetic field in a WSM produces a charge imbalance between
Weyl nodes of opposite chiralities. Such an imbalance gives rise to
a non-zero gyrotropic coefficient $\gamma$, a Hall-like contribution
to the dielectric tensor which determines the optical activity of
the system. A routine circular dichroism experiment can then potentially
detect the effect. We show how applying time-dependent $\boldsymbol{E}$
and $\boldsymbol{B}$ fields facilitates the isolation of the anomalous
contributions to the optical activity from possible non-anomalous
ones. Additionally, anomalous optical activity distinguishes between
Dirac and Weyl semimetals. In particular, Dirac semimetals do not
exhibit a chiral anomaly and thus, cannot develop an $\boldsymbol{E}\cdot\boldsymbol{B}$
induced valley imbalance. However, unlike in WSMs, even pre-existing
valley imbalances in Dirac semimetals do not contribute to $\gamma$
since Dirac nodes are achiral while $\gamma$ is directly sensitive
to the chirality of the system. We estimate the typical size of the
anomalous circular dichroism in WSMs and find it to be accessible
by current experiments. Finally, we point out that other experiments
that can measure material parameters that have the same symmetries
as the gyrotropic coefficient can also be used to probe the chiral
anomaly in WSMs as well as to distinguish them from Dirac semimetals.
\begin{acknowledgments}
We would like to thank the Packard Foundation for financial support.
\end{acknowledgments}
\bibliographystyle{apsrev4-1}
\bibliography{C:/Users/Pavan/Dropbox/CurrentProjects/references}

\end{document}